# The structure of the 3D - vortex lattices in microchip laser resonator.

*A.Yu.Okulov.*

*P.N.Lebedev Physical Institute of Russian Academy of Sciences*
*Leninsky prospect 53, 119991 Moscow, Russia , e_mail: okulov@sci.lebedev.ru*
*Present address: Department of Physics, University of Coimbra,*
*Coimbra 3000, Portugal : okulov@ci.uc.pt*

**Abstract.**

The interaction of the optical vortices leads to formation of spatially periodic structures of electromagnetic field which obey the scaling law typical to Talbot self-imaging.

___________________________________________________________________

Optical vortices are the natural component of spatiotemporal laser dynamics/1,2/. Their interaction leads to formation of spatially periodic structures of electromagnetic field which had been predicted via theoretical analysis/3,4/ and numerical modelling/1,5/. Recently the vortex lattices had been realized experimentally in broad area solid-state microchip lasers/2/. The stability and compact design of this devices provide definite advantages compared to **VCSEL** and $CO_2$ lasers in responsible applications such as atomic clocks etc.

The general theoretical approaches/3,4/ to spatially periodic structures based on necessary assumption of single frequency character of electromagnetic field structures show the experimental way for the selection of spontaneously formed arrays: the device cavity should be short enough, in order to guarantee the sufficiently large longitudinal mode separation interval $c/(2Ln)$ compared to gain linewidth/6/ ( *L* – the cavity length, *n* – refractive index, *c* – speed of light). This condition had been strictly fulfilled in experimental situation for **Nd:YAG** microchip cavity having length **2 mm,** where intermode interval was greater than **50 Ghz** /2/ (fig.1).

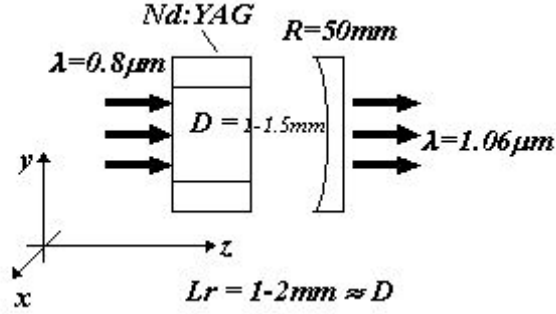

**Fig.1**

The scaling law for Fresnel number $N_{FR} = N_V^2$ /6/ was also fulfilled: the number of synchronised vortices $N_V$ was approximately equal to the square root from Fresnel number $N_{FR}$. The near-field intensity distributions show the smooth decreasing of the number of synchronised vortices along with the increasing of the cavity length /2, 6/.

In the present paper the efficient theoretical approach to this spontaneously formed structures is applied. We get spatially periodic solutions as a fixed points of nonlocal nonlinear maps in the form of convolution integral with nonlinear transfer function *f* describing amplification inside the gain slice /7/. The nonlinear integral equation for optical field takes into account the effect of spatial filtering of high spatial harmonics on transversely periodic gain grating. This form of evolution equation provides even more careful account of spatial filtering then Swift-Hohenberg equation /8/. We extract spatially periodic array of vortices from self-consistency condition in the framework of Fox-Lee approach /7/:

$$E_n(\vec{r}) = \iint f(E_n(\vec{r}\,')) \, K(\vec{r} - \vec{r}\,') \, d^2\vec{r}\,' \qquad (1)$$

$$K(\vec{r} - \vec{r}\,') = \frac{ik}{2\pi z} \exp(ik(\vec{r} - \vec{r}\,')^2 / 2z) \qquad (2)$$

The resulting plot of the transversal intensity and phase distributions in $z_T = 2\, p^2 / \lambda$ plane, i.e. at output mirror, are close to those, observed experimentally recently/2/. The visible rotation of intensity distribution on **45** degrees compared to phase distribution (fig.2) could be interpreted as interference of outer core of the vortex with his coherent neighbours. The maxima of intensity are in between zeros of amplitude, where the phase of the field is not determined.

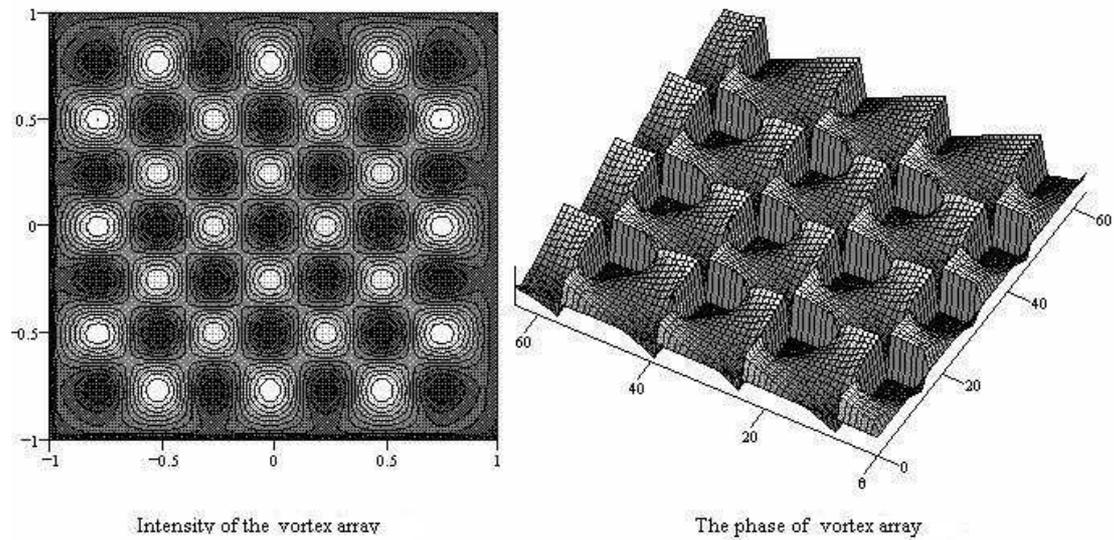

Intensity of the vortex array    The phase of vortex array

**Fig.2**

The main feature of these periodical structures is the coarse intensity distribution in the longitudinal direction (along the cavity axis *z*). In the middle of the cavity, i.e. at *z = 0*, the bright maxima form the grating with the same period **p** as on the mirrors (*z = -8, 8*), but this grating is shifted at **p/2** in transversal section. In the transversal sections corresponding to the quarter of Talbot distance $z_T = 2p^2/\lambda$ the period of the grating is divided by factor 2 (**p/2**) and intensity of this grating is smaller. Finally, because of periodicity of field distribution on one mirror, the field is self-imaged on the opposite mirror (fig.3):

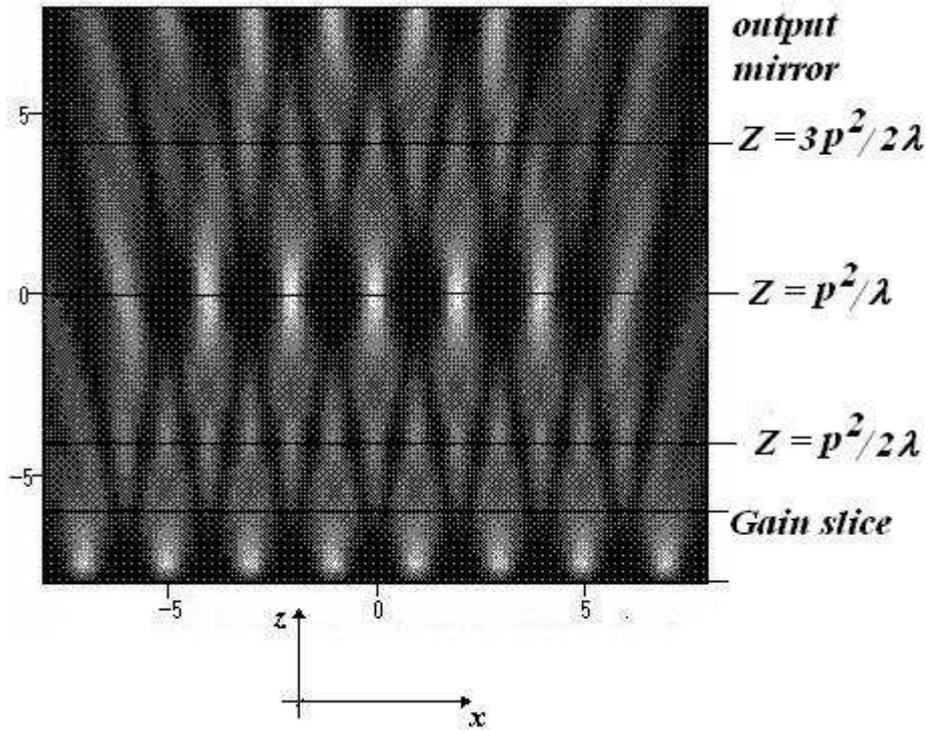

Transversal-longitudinal distribution of intensity of the 8-element array of the synchronized vortices.

**Fig. 3**

The theoretical model (1) works beyond the so-called mean field approximation, when electromagnetic field is assumed to be homogeneous along $z$ – axis/1/. The fine structure in fractional Talbot planes in fig.3 was obtained taking into account the propagation effects in paraxial approximation /6/.

These spatial structures are interesting from the point of view of compact integrated optical trap.

**References.**